\documentclass[twocolumn,superscriptaddress, amsmath, a4paper]{revtex4}

\usepackage{textcomp}
\usepackage{graphicx}
\usepackage[latin1]{inputenc}

\begin{document}


\title{Resolution in Focused Electron- and Ion-Beam Induced Chemical Vapor Deposition}

\author{Ivo Utke}
\email{ivo.utke@empa.ch}
\homepage{www.empa.ch}
\affiliation{Empa, Swiss Federal Laboratories for Materials Testing and Research, Materials- and Nanomechanics Laboratory, Feuerwerkerstr. 39, CH-3602 Thun, Switzerland}
\author{Vinzenz Friedli}
\affiliation{Empa, Swiss Federal Laboratories for Materials Testing and Research, Materials- and Nanomechanics Laboratory, Feuerwerkerstr. 39, CH-3602 Thun, Switzerland}
\affiliation{Advanced Photonics Laboratory, Ecole Polytechnique F\'ed\'erale de Lausanne (EPFL), CH-1015 Lausanne, Switzerland}
\author{Martin Purrucker}
\affiliation{Empa, Swiss Federal Laboratories for Materials Testing and Research, Materials- and Nanomechanics Laboratory, Feuerwerkerstr. 39, CH-3602 Thun, Switzerland}
\author{Johann Michler}
\affiliation{Empa, Swiss Federal Laboratories for Materials Testing and Research, Materials- and Nanomechanics Laboratory, Feuerwerkerstr. 39, CH-3602 Thun, Switzerland}

\date{\today}

\begin{abstract}
The key physical processes governing resolution of focused electron-beam and ion-beam-assisted chemical vapor deposition are analyzed via an adsorption rate model. We quantify for the first time how the balance of molecule depletion and replenishment determines the resolution inside the locally irradiated area. Scaling laws are derived relating the resolution of the deposits to molecule dissociation, surface diffusion, adsorption, and desorption. Supporting results from deposition experiments with a copper metalorganic precursor gas on a silicon substrate are presented and discussed.
\end{abstract}

\pacs{}

\maketitle


Local electron-beam induced chemical vapor deposition (CVD) is a physical phenomenon well known from the build up of carbon contamination since the beginning of electron microscopy \cite{Oatley1982}. Recent research has spent tremendous efforts on the systematic creation of functional nanoscale deposits by means of focused electron beams and - with the development of scanning ion microscopes - of focused ion beams. Organic, organo-metallic, and inorganic precursor gas molecules were supplied into the microscope chamber. Upon irradiation, deposition results from non-volatile dissociation products whereas etching occurs when a reaction of dissociation products with the substrate leads to the formation of volatile species. These local deposition and etching techniques have numerous potential applications in nanosciences including fabrication of attachments in mechanics \cite{Yu2000}, high-resolution sensors in magnetic, thermal, and optical scanning probe microscopy \cite{UtkeMFM2002, Edinger2001, Cast1999}, optical elements in nanooptics \cite{Sanchez2002, Nagata2005}, contacts in electronics \cite{Gopal2004}, and nanopores for ionic current measurements of cells and DNA in biology \cite{Nilsson2006, Danelon2006}. 

Few experiments analyze the physics of focused electron-beam (FEB) and focused ion-beam (FIB) induced deposition and etching. For FEB, Allen et al. \cite {Allen1988} noted that the spatial flux distribution of electrons passing through the adsorbed molecule layer consists of the incident primary beam (of up to several keV) and the emitted electrons: back scattered primaries and low-energy ($\leq 50~eV$) secondary electrons. The entire energy spectrum is responsible for the dissociation process. For FIB, Dubner et al. \cite{Dubner1998} showed that this spectrum is associated to the energy deposited into the substrate surface through the collision cascade generated by the primary ion beam.    

Our work is motivated by the fact that focused particle beam induced chemical vapor deposition and etching processes are widely used in nanoscale fabrication, but there are only very few attempts to describe the spatial resolution of this process theoretically. For a singular primary electron beam Silvis-Cividjian et al. \cite{Silvis-Cividjian2005} and Hagen et al. \cite{Hagen2006} concluded from Monte Carlo simulations that the ultimate resolution depends on the emitted secondary electron distribution. However, their assumption that the irradiated area is permanently covered with a monolayer of adsorbed molecules is idealized. In fact, as we will show, this coverage results from a balance of molecule depletion by dissociation and molecule replenishment strongly depending on adsorption, desorption, and diffusion. This is also White's et al. conclusion \cite{White2006} who studied the gas transport phenomena at the microscope chamber scale affecting the overall deposition rate. Müller's model \cite{Muller1971eng} for FEB induced deposition, which was later adapted by Haraichi et al. \cite {Hara1993} to gas-assisted ion-beam induced etching, takes the key processes of surface diffusion, dissociation, desorption, and adsorption into account via an adsorption rate equation. However, he used a flat-top beam distribution which allows no conclusions about resolution.

In this letter we present an adsorption rate model without the above mentioned limitations, considering two relevant peak distributions for the incident beam and for the emitted spectrum, and a non-dissociative Langmuir adsorption term. It allows to derive two scaling laws for resolution and to estimate important physical parameters of the process. Finally, we discuss results of carefully designed experiments, clearly supporting the theoretical conclusions. 

\begin{figure}
\includegraphics[width = 0.7\columnwidth]{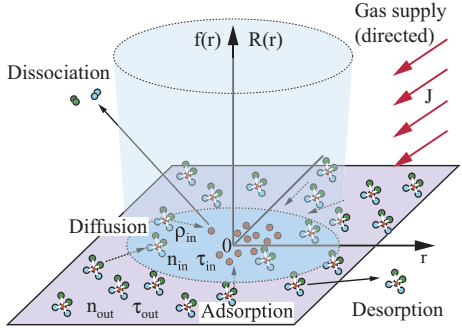}%
\caption{\label{fig:Process}
Reference system and schematics of processes involved in FEB induced deposition. Inside the irradiated area  precursor molecules are depleted by dissociation and replenished by adsorption and surface diffusion (dashed arrows). Symbols are defined in text.}
\end{figure}
The model assumes second-order kinetics of molecule dissociation by electrons. In a system with rotational symmetry the vertical FEB deposition or etch rate $R(r)$ (in units of dimension per unit time) as function of the distance $r$ from the centre of the primary electron (PE) beam is thus:
\begin{equation}
R(r) = Vn(r)\int_{0}^{E_{PE}}\sigma(E) f(E,r)dE \cong Vn(r)\sigma f(r)
\label{eqn:Rate}
\end{equation}

where $V$ is the volume of the decomposed molecule or etched atom, $n(r)$ is the number of adsorbed molecules per surface unit, $\sigma(E)$ is the energy dependent electron impact dissociation cross section, $E_{PE}$ is the energy of the PEs, and $f(E,r)$ describes the spatial flux distribution of the electron energy spectrum generated by the PEs. Since the energy integral can be solved only approximatively due to missing $\sigma(E)$ data of adsorbed molecules and uncertain parameter estimates entering in the Monte Carlo simulation of the emitted electron distribution \cite{Allen1988}, we use the simplified expression in Eq. \ref{eqn:Rate}, where $\sigma$ represents an integrated value over the energy spectrum. Such cross sections were measured for several relevant molecules \cite{Kohlmannvonplaten1992,Scheuer1986,Kunze1967,Petzold1991,Blauner1989,Lipp1996}. The spatial distribution $f(r)$ is a convolution of the Gaussian incident beam distribution $f(r) \propto exp(-r^2)$ with a peak function for the emitted spectrum which can be approximated by $f(r) \propto exp(-r)$. Full widths at half maximum (FWHM) of the emitted distributions range between $\sim 0.1~\text{nm (200 keV)}$ \cite{Hagen2006} and $2~\text{nm (1 keV)}$ \cite{Allen1988}. Similar considerations apply for FIB where the spatial distribution being responsible for molecule dissociation is a convolution of the primary beam distribution with the distribution of excited surface atoms generated by the collision cascade \cite{Dubner1998}. Table \ref{tab:Conditions} summarizes typical FWHM values of primary Gaussian beams.

Four key processes as depicted in Fig. \ref{fig:Process} are considered to determine the surface density $n(r)$ of adsorbed molecules: a) adsorption from the gas phase governed by the precursor flux $J$, the sticking probability $s$, and coverage $n/n_{0}$; b) surface diffusion from the surrounding area to the irradiated area governed by the diffusion coefficient $D$ and the concentration gradient; c) desorption of physisorbed molecules after a residence time $\tau$; d) molecule dissociation governed by the product $\sigma f(r)$. For the molecule adsorption rate $dn/dt$ follows:

\begin{eqnarray}
\frac{\partial n}{\partial t}= \underbrace{sJ\left(1-\frac{n}{n_0}\right)}_{\text{Adsorption}}+&&\underbrace{D \left(\frac{\partial^{2} n}{\partial r^{2}}+\frac{1}{r}\frac{\partial n}{\partial r}\right)}_{\text{Diffusion}}\nonumber\\
&&-\underbrace{\frac{n}{\tau}}_{\text{Desorption}}-\underbrace{\sigma f n}_{\text{Decomposition}}.
\label{eqn:AdsorptionRate}
\end{eqnarray}
The adsorption term in Eq. \ref{eqn:AdsorptionRate} describes a non-dissociative Langmuir adsorption, where $n_{0}$ is the maximum monolayer density given by the inverse of the molecule size. This adsorption type accounts for surface sites already occupied by non-dissociated precursor molecules and limits the coverage to $n_{0}$. All parameters other than $n=n(r,t)$ and $f=f(r,t)$ are considered constant.

Solving eq. \ref{eqn:AdsorptionRate} for steady-state ($dn/dt=0$) and neglecting the diffusion term we obtain $n(r)=sJ\tau_{\text{eff}}(r)$ with the effective residence time of the molecules $\tau_{\text{eff}}(r)=\left(sJ/n_{0}+1/\tau+\sigma f(r)\right)^{-1}$. The deposition or etch rate becomes

\begin{equation}
R(r)=sJ\tau_{\text{eff}} (r) V\sigma f(r)
\label{eqn:DepositionEtchRate}
\end{equation} 

and represents the deposit or etch shape at a given time. For any peak function $f(r)$ with a peak value $f_{0} = f(r=0)$, we can define the effective residence time in the center of the eletron beam $\tau_{\text{in}}=\tau_{\text{eff}}(0)=1/(sJ/n_{0}+1/\tau+\sigma f_{0})$ and the effective residence time far away from the electron beam center $\tau_{\text{out}}=\tau_{\text{eff}}(r \rightarrow \infty )=1/(sJ/n_{0}+1/\tau)$. The dimensionless ratio $\tilde\tau=\tau_{\text{out}}/\tau_{\text{in}} = 1 + \sigma f_{0} / (1/\tau + s J / n_{0})$ represents a measure for depletion of precursor molecules due to dissociation at the center of the beam. With a Gaussian beam $f(r) = f_{0} \exp({-r^{2}/2a^{2}})$ we derive the first scaling law of deposit size as function of depletion:

\begin{equation}
\tilde \varphi=\left\{\log_{2}(1+\tilde\tau)\right\}^{1/2}=\text{FWHM}_{\text{D}}/
{\text{FWHM}_{\text{B}}},
\label{eqn:FeatureSize}
\end{equation}

where $\text{FWHM}_{\text{B}}$ and $\text{FWHM}_{\text{D}}$ are the full widths at half maximum of $f(r)$ and $R(r)$ or, in other words, the FWHMs of the incident beam and the deposit. The idealized case of zero depletion, i.e. permanent monolayer coverage, corresponds to $\tilde\tau=1$. Then deposition or etching proceeds in the electron-limited regime and the deposit (or etch) size corresponds to the electron beam distribution since the logarithmic term becomes 1. With increasing depletion the deposit (or etch) size becomes steadily larger than the beam size. For the peak function $f(r) \propto exp(-r)$ the expononent in eq. \ref{eqn:FeatureSize} becomes 1. The degree of depletion strongly depends on the dissociation frequencies $\sigma f_{0}$ summarized for FEB and FIB in Table \ref{tab:Conditions}. In order to replenish the dissociated molecules inside the continuously irradiated area by gas transport only, we need $\tilde \tau \rightarrow 1$, i.e. the precursor supply frequency $s J / n_{0}$ should exceed $\sigma  f_{0}$, being equivalent to $>2\times10^{3} \text{ML}/s$ (monolayers per second) for FEB. This corresponds to a precursor flux on the substrate of $J = 2\times 10^{17}~\text{molecules}~\text{cm}^{-2}~\text{s}^{-1}$, setting $s = 1$ and taking $n_{0} = 10^{14}~\text{cm}^{-2}$ as typical value. For FIB several orders of magnitude larger gas supply would be needed. Desorption frequencies are situated around $\tau^{-1} = 10^{0}\dots10^{3}~\text{Hz}$ \cite{Edinger2000, Scheuer1986}. Above estimations clearly suggest that most of the FEB and FIB processing experiments were performed in the precursor-limited regime limiting the minimum deposit or etch size.

\begin{table}
\caption{\label{tab:Conditions}
Typical ranges of incident peak flux $f_{0}$ and size $\text{FWHM}_{\text{B}}$ of a focused electron beam (5 keV, field emission) and an ion beam (30kV, Ga$^{+}$). Representative ranges for $\sigma$ were collected from \cite{Kohlmannvonplaten1992,Scheuer1986,Kunze1967} for FEB and from \cite{Petzold1991,Blauner1989, Lipp1996} for FIB.}
\begin{ruledtabular}
\begin{tabular}{ccccc}
 & $f_0$ & $\text{FWHM}_{\text{B}}$ & $\sigma$  & $\sigma f_{0}$ \\
 & [1/nm$^2$s] & [nm] & [nm$^{2}$] & [1/s] \\
\hline
FEB & $8 \times 10^{6} \dots$ & $2.5 \dots$ & $2 \times10^{-4} \dots$ & $2 \times 10^{3} \dots$ \\
  & $5 \times 10^{7}$ & $100$ & $2 \times 10^{-1}$ & $1 \times 10^{7}$ \\
\hline
FIB & $2 \times 10^{5} \dots$ & $7 \dots$ & $10\dots$ & $2 \times 10^{6} \dots$ \\
  & $5 \times 10^{6}$ & $100$ & 50 & $2.5 \times 10^{8}$ \\
\end{tabular}
\end{ruledtabular}
\end{table}

\begin{figure}
\includegraphics[width = 1\columnwidth]{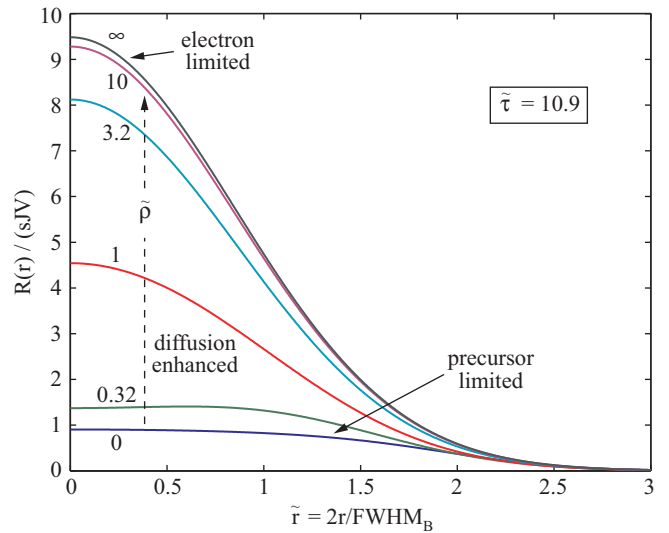}%
\caption{\label{fig:Diffusion}
Normalised steady-state deposition rate at indicated depletion from Eq. \ref{eqn:Rate} representing the deposit shape. The  diffusive replenishment $\tilde \rho = 2\rho_{\text{in}} / \text{FWHM}_{\text{B}}$ is varied. Note the shape transition from flat top, $\tilde \rho = 0$, indented, $\tilde \rho = 0.32$, to Gaussian, $\tilde \rho = \infty$.}
\end{figure}

In the following we show under which conditions substantial replenishment by surface diffusion can be expected. Solving equation \ref{eqn:AdsorptionRate} numerically with MATLAB for steady-state ($dn/dt=0$), the boundary conditions $n(r \rightarrow \infty) = n_{\text{out}}=sJ\tau_{\text{out}}$ and $dn(r = 0)/dr = 0$, and with a Gaussian distribution $f(r) = f_{0} \exp({-r^{2}/2a^{2}})$ we get $n(r)$ and finally $R(r)$. A plot against the dimensionless variable $\tilde r = 2r/\text{FWHM}_{\text{B}}$ for a given depletion and diffusive replenishment results in a unified representation of deposit shapes for any $\text{FWHM}_{\text{B}}$, see Fig. \ref{fig:Diffusion}. Replenishment by diffusion is described by the dimensionless ratio $\tilde \rho = 2\rho_{\text{in}}/\text{FWHM}_{\text{B}}$, relating the diffusion path in the center of the beam $\rho_{\text{in}} = (D \tau_{\text{in}})^{1/2}$ with D being the diffusion coefficient, to the beam size. For $\tilde \rho = 0$, the flat top shape is defined by Eq. \ref{eqn:DepositionEtchRate}. With increasing diffusive replenishment deposits change into indented and round apex shapes, since adsorbed molecules increasingly reach the centre of the irradiated area before being dissociated. Hence both deposition rate \textit{and} resolution increase. The maximum diffusion enhancement in deposition rate becomes $R(\tilde \rho=\infty)/R(\tilde \rho=0) = \tilde \tau $ at $r=0$. For $\tilde \rho \rightarrow \infty$, Eq. \ref{eqn:DepositionEtchRate} simplifies to $R(r) = n_{\text{out}} V \sigma  f(r)$ since any depletion is entirely compensated by diffusion and a permanent monolayer coverage provided. In other words, the electron-limited regime is established and the deposit shape corresponds to the electron beam distribution $f(r)$.

\begin{figure}
\includegraphics[width = 1\columnwidth]{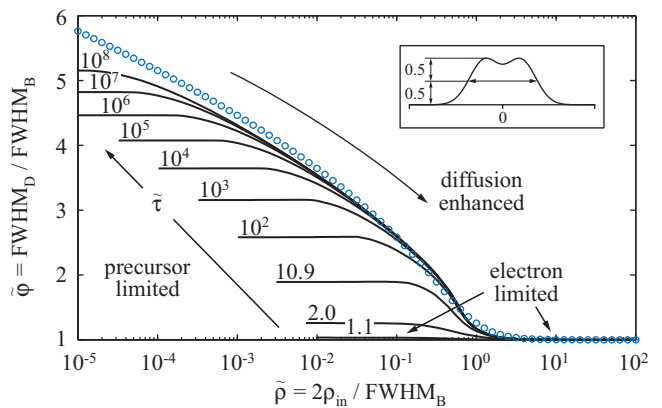}%
\caption{\label{fig:FeatureSize}
Normalized deposit size vs. normalized diffusion path for varying depletion (indicated). At $\tilde \rho = 2$ the diffusion path equals the beam size. The curve for $\tilde \tau = 10.9$ corresponds to the FWHMs of shapes in Fig. \ref{fig:Diffusion}. Circles represent Eq. \ref{eqn:FeatureSizeDiff}. The inset shows the  $\text{FWHM}_{\text{D}}$ definition of indented deposits.}
\end{figure}

Figure \ref{fig:FeatureSize} represents a graph relating the dimensioneless deposit size $\tilde \varphi = \text{FWHM}_{\text{D}} / \text{FWHM}_{\text{B}}$ to irradiative depletion and diffusive replenishment. For any depletion with $\tilde \rho \geq 2$ we get $\tilde \varphi \leq 1.03$, i.e. the deposit size is within 3\% close to $\text{FWHM}_{\text{B}}$ when the related diffusion path inside the irradiated area becomes at least comparable to the size of the electron beam distribution. The deposits become broader than the electron beam for $\tilde \tau > 1$ and small $\tilde \rho$, branching out into constant maximum size given by Eq. \ref{eqn:FeatureSize} at negligible diffusive replenishment $\tilde \rho \rightarrow 0$. Figure 3 holds independently of how diffusive replenishment is experimentally achieved: either via the beam size $\text{FWHM}_{\text{B}}$ (using the focus of the beam) or via the diffusion path $\rho_{\text{in}}$ (changing precursor diffusion). The second scaling law of deposit size as function of diffusive replenishment is obtained as (circles in figure \ref{fig:FeatureSize}):

\begin{equation}
\\ \tilde \varphi \approx \left\{ \log_{2}\left(2 + \tilde \rho^{-2} \right) \right\}^{1/2}.
\label{eqn:FeatureSizeDiff}
\end{equation}

For the exponential peak function $f(r) \propto exp(-r)$ the expononent in Eq. \ref{eqn:FeatureSizeDiff} becomes again 1. Both Eq. \ref{eqn:FeatureSize} and Eq. \ref{eqn:FeatureSizeDiff} give upper limits for $\tilde \varphi (\tilde \tau)$ and $\tilde \varphi (\tilde \rho)$. The smaller value of both defines the minimum deposit (or etch) size with respect to the beam size. 

Now we present FEB deposit shape measurements obtained with a special designed SEM compatible atomic force microscope (AFM). The advantage is that AFM reveals three-dimensional topography features, like indented shapes, which are difficult to resolve in SEMs due to edge contrast effects. We used Cu(II)-hexafluoroacetylacetonate precursor molecules impinging on a native Silicon substrate with $J/n_{0}=10 \text{ML/s}$ under irradiation with a 5 keV Gaussian electron beam ($f_{0}= 9 \times 10^{4}~  \text{nm}^{-2}\text{s}^{-1}$ and $\text{FWHM}_{\text{B}}=110 ~\text{nm}$). An indented shape with $\text{FWHM}_{\text{D}}=200~\text{nm}$ is observed, hence $\tilde \varphi = 1.8$. Assuming $\tau = 10^{-3}~\text {s}$ and $s = 1$ we get from Eq. \ref{eqn:FeatureSize} $ \tilde\tau = 8.9$, i.e. $\sigma  \sim 0.09  ~\text{nm}^{2}$. From Eq. \ref{eqn:FeatureSizeDiff} follows $\tilde \rho = 0.37$, i.e. $\rho_{\text{in}} = 20~\text{nm}$. Using the relation $\rho_{\text{in}} \simeq (D / \sigma  f_{0})^{1/2}$ for  $\tilde \tau \gg 1$ results in $D \sim 3\times 10^{-8} ~\text{cm}^{2} \text{s}^{-1}$. 

\begin{figure}
\includegraphics[width = 1\columnwidth]{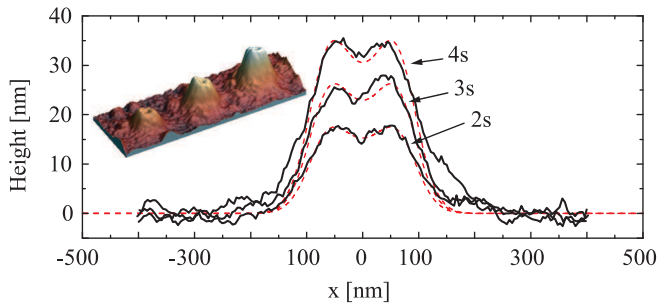}
\caption{\label{fig:AFM}
AFM image and line scans of FEB deposits from Cu(hfac)$_2$ precursor. Exposure times are indicated. The indented apex shapes are due to depletion. Dashed lines represent fits of Eqn. \ref{eqn:Rate}.}
\end{figure}

The values for the depletion, cross section, and diffusion coefficient represent lower limit estimations since the same FWHM-ratio $\tilde \varphi$ can be obtained with larger depletion \textit{and} larger diffusive replenishment, see figure \ref{fig:FeatureSize}. Taking as maximum dissociation cross section the molecule size $\sigma  = 0.6~\text{nm}^{2}$ \cite{Gromilov2004}, we get $\tilde \tau = 60$. The corresponding upper limit estimate for the diffusion coefficient is derived from the shape fit in Fig. \ref{fig:AFM} to be $D = 4\times 10^{-7} \text{cm}^{2} \text{s}^{-1}$. 
Finally, we estimate diffusion coefficients needed for establishing the electron-limited regime under typical irradiation conditions summarized in table \ref{tab:Conditions}. Compensation of depletion by surface diffusion requires a beam focus smaller than the molecule diffusion path, $ \tilde \rho \geq 2$, see figure \ref{fig:FeatureSize}. Together with the relation $\rho_{\text{in}} \simeq (D / \sigma  f_{0})^{1/2}$ for $\tilde \tau \gg 1$ we obtain $D = 10^{-12}\dots10^{-6}~\text{cm}^{2}~\text{s}^{-1}$ for FEB and $D = 10^{-7}\dots10^{-3}~\text{cm}^{2}~\text{s}^{-1}$ for FIB.

In conclusion, we quantified the crucial role of depletion and diffusive replenishment on deposit resolution for two relevant distributions: an incident Gaussian beam and an emitted distribution with exponential decay via an adsorption rate model. Our model is applicable to gas-assisted deposition and etching with focused electron- and ion beams and thus covers numerous applications in fundamental research and in nano-scale fabrication. We demonstrated how physical parameters can be determined from  fitting experimental deposit shapes with our model. An extension of the studies to different beam shapes or experimental arrangements like pulsed beams at different temperatures is straightforward. It will enable the systematic determination of all physical key parameters involved in the process thus opening the door to the controlled fabrication of tailored nanoscale devices by charged particle beam induced CVD and etching. 

We acknowledge financial support from the European Commission, FP6 Integrated Project NanoHand
(IST-5-034274).


\end{document}